\documentclass{epl}

\shorttitle{Diverging length scale in the {\sc mct} of glasses}
\title{Diverging length scale and upper critical dimension 
in the Mode-Coupling Theory of the 
glass transition}
\author{Giulio Biroli\inst{1}   
	\And Jean-Philippe Bouchaud\inst{2}}
\institute{
\inst{1} Service de Physique  Th{\'e}orique Centre d'{\'e}tudes de Saclay \\ 
  Orme des
Merisiers, 91191 Gif-sur-Yvette Cedex, France\\
\inst{2} Service de Physique de l'{\'E}tat
Condens{\'e},
 	       Centre d'{\'e}tudes de Saclay \\ 
  Orme des
Merisiers, 91191 Gif-sur-Yvette Cedex, France}

\pacs{05.70.Jk}{Critical point phenomena}
\pacs{64.60.Ht}{Dynamic critical phenomena}
\pacs{64.70.Pf}{Glass transitions}

\begin{document}
\newcommand \be  {\begin{equation}}
\newcommand \bea {\begin{eqnarray} \nonumber }
\newcommand \ee  {\end{equation}}
\newcommand \eea {\end{eqnarray}}
\newcommand{\siml}{\stackrel{<}{\sim}}
\newcommand{\subs}[1
]{{\mbox{\scriptsize #1}}}
\maketitle

\begin{abstract}
We show that the glass transition predicted by the Mode-Coupling Theory
({\sc mct}) is a critical phenomenon with a 
diverging length and time scale associated to the cooperativity 
of the dynamics. We obtain the scaling exponents $\nu$ and $z$ 
that relate space and time scales to the distance from criticality, 
as well as the scaling form of
the critical four-point correlation function. 
However, both these predictions and other well known {\sc mct} results are {\it 
mean-field} in nature and are thus expected to change below 
the upper critical dimension $d_c=6$, as suggested by different forms of the Ginzburg 
criterion.
\end{abstract}

One of the most striking property of glassy materials is the extremely fast rise of their relaxation
time (or viscosity) as the temperature is lowered or the density increased \cite{Tarjus}. 
The basic mechanism for 
this spectacular slowing down is not well understood, but it is reasonable to think that it is intimately
related to cooperative effects. The dynamics becomes sluggish because larger and larger regions of the
material have to move simultaneously to allow for a substantial motion of individual particles. Although 
this qualitative idea has pervaded the glass literature for many years \cite{Tarjus}, it is only quite recently that
a proper measure of cooperativity (and of the size of the rearranging regions) was proposed and measured 
experimentally \cite{Ediger} and in
numerical simulations \cite{Harrowell,Onuki,Glotzer}. The idea 
is to measure how the `unlocking' events are correlated in space; technically,
this involves a four-point density correlation function (see below)
from which one can extract a growing dynamical correlation length
\cite{Onuki,Glotzer,Berthier1}. 
Recent extensive numerical simulations in Lennard-Jones 
systems have confirmed the crucial importance of this growing length
scale for the dynamics of the system \cite{Onuki,Glotzer,Berthier1}. 
Furthermore, the four-point correlation function is found to have 
scaling properties similar to those expected near a critical point
\cite{Onuki,Glotzer,Berthier2}, 
suggesting that the physics of the glass transition 
should be understood as a critical phenomenon 
dominated by large scale fluctuations. This ingredient in fact appears in various forms 
in several recent phenomenological approaches
\cite{Wolynes,Tarjus2,Chandler,Berthier2}. This is,
at first sight, in plain contradiction 
with the Mode-Coupling Theory ({\sc mct}) 
of the glass transition. {\sc mct} is considered 
by many to be the closest to a first principle, 
microscopic theory of glasses yet achieved, with many qualitative and 
quantitative successes in explaining various experimental and 
numerical results \cite{Gotzeoriginal,Gotze1,Cates}. 
However, freezing in {\sc mct} was argued 
repeatedly by its founders to be a small scale phenomenon, 
the self-consistent blocking of the particles in their 
local cages. Since no small wavevector singularities seem to
exist in {\sc mct}, power-laws and scaling are only expected in time
but not in space \cite{Gotze1}. This is surprising since on general grounds 
a diverging relaxation time can only arise from processes
involving an infinite number of particles.
This point of view was challenged by Franz and Parisi \cite{FP} 
in the context of the so-called `schematic' {\sc mct} 
(see also \cite{Wolynes} for an early insight). 
This simplified version of the theory  
is (formally) equivalent to describing mean-field spin glasses 
with three body interactions, for which the physics 
is known in great details. At the (ergodicity breaking) 
critical temperature $T_c$, the curvature of the 
relevant TAP states is known 
to vanish \cite{Barrat}. Therefore, one expects, and indeed finds, 
that a susceptibility diverges when $T \to T_c$. 
This susceptibility turns out to
be the precise analogue, in the context of this spin model, 
of the four-point correlation function mentioned above. 
By analogy with usual second order phase transitions, the analysis of
\cite{Wolynes,FP} suggests that the 
{\sc mct} freezing transition is in fact accompanied by the divergence 
of the correlation length of the four-point correlation function. 
 
\begin{figure}
\onefigure[width=6.5cm]{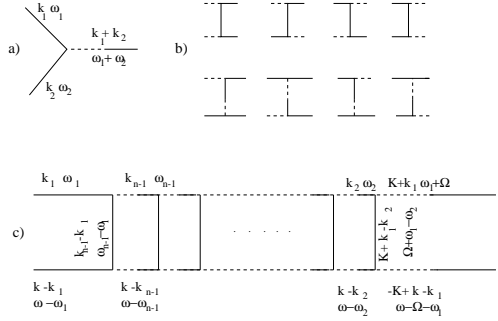}
\caption{Basic vertex of the {\sc mct} (a), and a ladder diagram that contribute to the
divergence of the four-point susceptibility (c). Other ladders can be obtained by 
combining the eight elementary blocks shown in (b). The full lines represent correlation
functions, whereas the mixed dashed-full lines are response functions.}
\label{fig1}
\end{figure}

In this letter we show that the {\sc mct} dynamical transition in finite 
dimensions must indeed be understood as a critical phenomenon: 
dynamical correlations become long-range both {\it in time and space}, 
in agreement with the insight of \cite{Wolynes,FP}.  
We obtain the {\sc mct} dynamical scaling exponents $\nu$ and $z$ 
that relate space and time scales to $|T-T_c|$, as well as the scaling form of
the critical four-point correlation function. 
However, these results (as well as all other quantitative {\sc mct} predictions) are 
{\it mean field} in nature and change in dimensions less than $d_c=6$ due to 
long-wavelength fluctuations, as confirmed by
different forms of the Ginzburg criterion.
Our strategy is similar to the one used for ordinary critical phenomena. 
Consider the ferromagnetic Ising transition as an example. In that
case one can show
that there exists a certain functional of the magnetisation field such that (a) 
its first derivative leads to exact equations for the magnetization and (b) its 
second derivative is the inverse of the spin-spin correlation function. 
In general one
cannot compute this functional exactly but one can guess its form using symmetry 
arguments, or compute it approximately in a diagrammatic
expansion. Its simplest version corresponds to the Ginzburg-Landau
free energy functional. 
The saddle point equations for the magnetization then leads to the mean field 
description of the transition. One finds in particular that the singular behaviour
at the transition is related to the vanishing of a `mass'. This has 
two important implications: 
(1) the spin susceptibility diverges at the transition, (2) because
of the vanishing mass 
the corrections to the mean field result computed by adding more diagrams blow up 
whenever $d<d_{c}=4$, i.e. spatial fluctuations change the critical behavior.
We shall show that exactly the same scenario takes place within the {\sc mct} of the glass 
transition, except that the order parameter is now a two point function, the 
dynamical density-density correlation function. 
The analogue of the spin-spin correlation function is
therefore the following four point correlation:
\begin{equation}
G_4(\vec r,t; \vec \delta, \tau) = \langle \rho(0,0)  \rho(\vec{\delta},\tau) 
\rho(\vec r,t) 
\rho(\vec r+\vec \delta,t+\tau) \rangle-\langle \rho(0,0)  
\rho(\vec{\delta},\tau) \rangle \langle
\rho(\vec r,t) 
\rho(\vec r+\vec \delta,t+\tau) \rangle,
\end{equation}
where $\rho(\vec{x},t)$ is the density fluctuation at position $\vec x$ and time $t$
and $\langle \cdot \rangle$ denotes an average over the dynamics.
Its intuitive interpretation (for example in the case where 
$\vec{\delta}=0$) is as follows:
if at point $0$ an event has occurred that 
leads to a decorrelation of the local density over the time scale $\tau$, 
what is the probability 
that a similar event has occurred a distance 
$\vec r$ away, within the same time interval $\tau$,
but shifted by $t$? In other words, $G_4(\vec r,t;\vec{\delta},\tau)$ measures 
the cooperativity of the dynamics. 
Different values of $\vec r$ and $t$ allow one to measure the
full space-time structure of dynamical cooperativity, and are the analogue of the space and 
time separation entering the standard spin-spin correlation function
in critical dynamics. The quantities $\vec \delta$ and $\tau$,
on the other hand, serve to define the `order parameter', 
i.e. the density-density correlation 
function $C(\vec{\delta},\tau )=\langle \rho(0,0)
\rho(\vec{\delta},\tau) \rangle$. Previous studies have
focused on the case $t=0$, and ${\delta}$ smaller than the
particle radius, to which one can associate a `susceptibility' 
$\chi_4(\tau)$ by integrating over space.
More generally,
we shall focus on a wave-vector dependent, a.c. susceptibility as:
\be
\chi_4(\vec k,\omega; \vec{K},\Omega)= 
\int d^d \vec r \, d^d \vec \delta \, 
dt \,d\tau \, e^{-i \vec k \cdot \vec r- i \omega t -i
\vec{K} \cdot \vec{\delta}-i \Omega \tau} 
\, G_4(\vec r,t; \vec{\delta}, \tau).
\ee
The quantities $G_{4}$ or $\chi_4$ can be computed as in the ferromagnetic case
inverting the second derivative of an appropriate functional.
Within any field theoretical derivation of {\sc mct} (e.g. the 
Das and Mazenko formulation \cite{DM} or more heuristic derivations 
\cite{Kawasaki,Cates}), the functional alluded to above can be constructed as \cite{cirano}:
\begin{equation}\label{2PI}
F (G)=-\frac{1}{2}\mbox{Tr} \log G+\frac{1}{2}\mbox{Tr}\, G_{0}^{-1} G+\Phi_{2PI} (G)
\end{equation}
where $G$ and $G_{0}$ are compact notations for the full and the bare
propagator of the theory, and $\Phi_{2PI} (G)$ is the sum of all two
particle irreducible Feynman diagrams (that cannot be decomposed in two
disjoint pieces by cutting two lines) constructed with the vertices of
the theory and using the full propagator as line. 
The first derivative of $F$ gives back the exact equations 
on $G$, whereas the four point function is obtained inverting its second derivative, 
which is nothing else than $G^{-1}G^{-1}$ (coming from
the derivative of the first term in (\ref{2PI})) 
minus the derivative of the self-energy with
respect to the two point function.
Thus, since {\sc mct} is tantamount to 
only retaining the `bubble' diagram, the 
four point functions are obtained from ladder diagrams, such
as the one drawn in Fig. \ref{fig1}\footnote{From a more general point 
of view, one can obtain all the MCT predictions using a Landau 
expansion of $F(G)$ in $G(t,T)-G(t=\infty ,T_{c})$,
justified in dimension larger than six. 
This derivation, which unveils the generality of the
MCT predictions, will the subject of a future publication.}. 

In order to understand the mechanism that leads to a divergence of the 
four-point correlation, consider the ladder diagram shown in Fig. 1. The $n^{th}$ order 
contribution to $\chi_4(\vec k,\omega,\vec{K},\Omega)$ reads:
\bea
& &\Delta_n \chi_4(\vec k,\omega,\vec{K},\Omega)=
\int d^d \vec k_1 d^d \vec k_2 ...d^d \vec k_{n-1} 
\int d\omega_1 d\omega_2 ... d\omega_{n-1}
C(\vec k_1,\omega_1) C(\vec k-\vec k_1,\omega-\omega_1) \times\\ \nonumber
& &{\cal M}_{\vec k,\omega}(\vec k_1,\vec k_{n-1};\omega_1,\omega_{n-1})
{\cal M}_{\vec k,\omega}(\vec k_{n-1},\vec k_{n-2};\omega_{n-1},\omega_{n-2})
....{\cal M}_{\vec k,\omega}(\vec k_{2},\vec k_{1}+\vec{K};\omega_{2},\omega_{1}+
\Omega)
\eea
where the matrix ${\cal M}$ is defined as:
\[
{\cal M}_{\vec k,\omega}(\vec k_{2},\vec k_{1};\omega_{2},\omega_{1})=T^{2}
\frac{R(\vec k_1,\omega_1)}{\rho k_{1}^{2}} \frac{R(\vec k-\vec
k_1,\omega -\omega_1)}{\rho (k-k_{1})^{2}} 
C(\vec k_2 - \vec k_1, \omega_2 -\omega_1)
V(\vec k_1, \vec k_{2} , \vec k),
\]
where 
$\rho$ is the particle density and $C$ is the correlation function and $R$ the response 
function (assumed to be related to $C$ by the fluctuation-dissipation theorem
\footnote{This property is certainly true for the $p=3$ spin-glass for $T > T_c$ 
but may be problematic for {\sc mct}. However, we believe that the following 
conclusions are independent of the strict validity of {\sc fdt}.}) for the density in 
Fourier space, and 
\[
V= \{\vec k_{1}\cdot [\vec k_{2}c (\vec k_{2})+ (\vec k_{1}- \vec k_{2})c
(\vec k_{1}- \vec k_{2})]\}\{
(\vec k_{1}-\vec k)\cdot [(\vec k_{2}-\vec k)c (\vec k_{2}-\vec k)+
(\vec k_{1}-\vec k_{2})c (\vec k_{1}-\vec k_{2})] \}
\]
where $c(\vec k)$ is the direct correlation function.
If one follows the so-called schematic approximation \cite{Gotze1,Gotzeoriginal}
where all structure factors $S(\vec k)$ appearing inside integrals are approximated by $A\delta (k-k_{0})$ and all ingoing and outgoing momenta 
have modulus $k_{0}$, then all $k$-dependence disappears. Replacing the resulting
($k$-independent) four leg vertex $T^{2}\hat V=T^{2}S (k_{0}) k_{0}A^{2}/ (4 \pi^{2}
\rho)$ with $T^{2}\hat V_{3-spin}=3$ (where $T$ is the temperature), 
one gets exactly the same diagrams of the mean field $p=3$ case.  
For $T < T_c$ and for small frequencies, the correlation 
function acquires a
(non-ergodic) contribution: $C(\omega)=f\delta(\omega) + C_\subs{reg}(\omega)$ 
(where $f$ is usually denoted 
$q$ in the spin-glass literature), while the response function tends 
to a constant at low frequencies given by $(1-f)/T$ 
(with power-law corrections in $\omega$, see 
below). Therefore, one obtains, as the dominant contribution: $
\Delta_n \chi_4(\omega,\Omega) = C(\Omega) \delta(\omega) f  
[\hat V (1-f)^2 f]^{(n-1)}
$.
Clearly, the series diverges when $(1-f)^2 f \hat V= 1$, an 
equation precisely satisfied at 
the mode coupling temperature $T_c$, and signaling criticality 
(see, e.g. \cite{Gotze1,Gotzeoriginal,Barrat}). 
Hence, the asymptotic value of 
$\chi_4(\tau \to \infty)$ is divergent at $T_c$. 
One can show, using a transfer matrix method to analyze 
all diagrams generated from the eight 
building blocks drawn in Fig. 1, that the above singularity is unchanged, i.e. $\chi_4$
indeed diverges as $(1-(1-f)^2 f \hat V)^{-1}$ \footnote{Actually a
complete calculation should also take into account the diagrams with extra vertices coming
from imposing an initial condition at equilibrium. Although these
diagrams do not change the singular behavior for $\tau \rightarrow
\infty $ they are responsible for the vanishing of $\chi _4$ when $t \to \infty$.}.
 Since $f - f_c \sim \sqrt{\epsilon}$ for $T < T_c$, we conclude that $\chi_4$ 
diverges as $\epsilon^{-1/2}$ for $T \nearrow T_c^-$, indeed in 
agreement with the results of \cite{FP}. 

Our aim in the following is to understand in details how this divergence is affected 
by non zero frequencies and wave-vectors. In order to do
so, we rely on the detailed results known about $C$ and $R$ in the vicinity of $T_c$. 
One knows 
for example that when $T = T_c - \epsilon$, the plateau in $C$ is reached after a time 
$\tau_f \sim \epsilon^{-1/2a}$, where $a$ is a non-trivial exponent. 
In the regime $\tau_f^{-1} \ll \omega \ll 1$, the response function acquires an extra 
contribution proportional 
to $\omega^a$ \cite{Gotzeoriginal}. On the other hand, when 
$T = T_c + \epsilon$, two time scales diverge 
with different 
exponents. One is again the time $\tau_f$ to reach the (pseudo-)plateau 
value $f$ of 
the correlation function, and the second is the {\it terminal} time $\tau_t 
\sim \epsilon^{-\gamma}$ (with $\gamma=1/2a+1/2b$),
beyond which the correlation finally drops to zero. Both the exponents $a$ and $b$ are non universal; for example,
in the mean-field $p=3$ case one has $a \approx 0.395$ and $b=1$. 
Using these results for $T < T_c$ and for non zero $\omega$ 
(corresponding to the shift in the measurement times for 
the four-point correlation function) we find that the term 
$1-(1-f)^2 f \hat V\sim \sqrt{\epsilon}$ leading to 
the singularity, is replaced by $\sqrt{\epsilon}+Z 
\omega^a$, where $Z$ is a constant.
Thus, the only characteristic 
time scale is, for $T < T_c$, the plateau time $\tau_f$. 
Resumming the series $\sum_n\Delta_{n}\chi_{4}$ is much more subtle in the case $T > T_c$.
A naive transposition of the results for $T<T_{c}$ to $T>T_{c}$ suggests 
the following: since the $1/\sqrt{\epsilon}$ divergence of $\chi_4$ for $T<T_{c}$ 
can be traced back to $f_c -f \sim \sqrt{\epsilon}$, then the divergence  for $T>T_{c}$
should be $1/\epsilon$ because now $f-f_{c}\simeq \epsilon$.
Furthermore, from the form of the ladder diagrams one expects
that both the correlation time scale in the $t$ direction and the 
peak time $t^*$ of $\chi_4(\tau)$ are
given by the terminal time scale $\tau_t$. 
The above predictions turn out to be in perfect 
agreement with the numerical solution 
of the $p=3$ mean-field case shown in Fig. 2 
\footnote{Note that these numerical results 
do not coincide with the results reported in \cite{FP}, 
obtained with the same code; in particular, the height 
of the peak was found to diverge as $\epsilon^{-1/2}$ and 
not as $\epsilon^{-1}$. A possibility is that the 
`conjugated field' used in \cite{FP} was not small enough.},
and can in fact be obtained analytically using another route \cite{BBB}.   
\begin{figure}
\onefigure[width=7cm,angle=270]{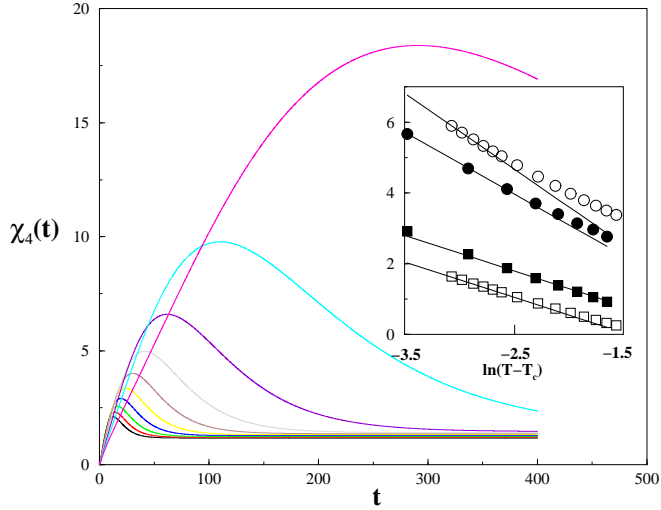}
\caption{Four-point susceptibility $\chi_4(\tau)$ in the mean-field $p=3$ spin-glass,
obtained by integrating the dynamical equations obtained in \cite{FP},
for various temperatures $T > T_c$ ($T_c=0.612$). The inset shows, in
log-log scale, the 
position $t^*$ (black circles) and the height $\chi_4(t^*)$ (black squares) of the peak as a function of $T-T_c$, and the plain lines are power-laws with exponents $-1$ and $-(1+a)/2a=-1.76$, corresponding to the `naive' argument discussed in the text.
We confirmed these results by also solving numerically the mixed $p=3+6$ case, for which 
$a=0.345$ and $b=0.717$ (open symbols). These last curves have been shifted
for clarity. 
 }
\label{fig2}
\end{figure}

Returning to {\sc mct} in finite dimensions, one has to include the wavevector dependence of
$C$, $R$ and $V$. Assuming ergodicity is broken, $\chi_4$ 
for $\vec k=0$ is found to diverge when the largest eigenvalue of matrix $M$ is
equal to unity:
\be\label{critical}
M(\vec k_2,\vec k_1)=  \frac{\rho}{k_{2}^{4}}S (\vec k_{1})S (\vec k_{2})S
(\vec k_{1}-\vec k_{2}) V(\vec k_1, \vec k_2, \vec{0}) (1- f_{ k_1})^2 f_{\vec k_2 - \vec k_1}
\ee
with, for $T \leq T_c$: $f_{ q}/ 
(1-f_{ q}) = \frac{\rho}{2 q^{4}}\int 
{d^{3}k'}/{(2\pi)^{3}} S
(\vec q)S (\vec k')S (\vec q-\vec k')  V(\vec{q},\vec{k'}, \vec{0})f_{k}f_{\vec q-\vec k'}$. 
The spectrum of $M$ has been previously studied in detail because it controls the
convergence of the above equation on $f_{q}$ when solved by an iterative procedure.
It has been established (see \cite{Gotze1}) that the maximum eigenvalue of $M$ 
is non degenerate and approaches one as $\sqrt{\epsilon}$ 
at the
transition, which in turn leads to the famous {\sc mct} singularity of $f - f_c$. 
The effect of a non zero wave-vector $\vec k$ can be understood using
perturbation theory. By symmetry, it is clear that the correction to
the largest eigenvalue 
must be, for small $\vec k$, of order $k^2$. This leads for small $k$ and
$T<T_{c}$ to a propagator  $(\Gamma k^2 + \sqrt{\epsilon})^{-1}$. 
Naively transposing the above results
for $T > T_c$ gives a propagator behaving as $(\Gamma' k^2 + \epsilon)^{-1}$. 
Thus, we conclude that the four-point
correlation computed, say, for $t=0$ and $\vec \delta$ of the order of 
the particle radius, and $\tau$ of the order of the
correlation time scale ($\tau_{f}$ for $T<T_{c}$ and $\tau_{t}$ for
$T>T_{c}$) will behave as the two-point correlation in standard critical phenomena, 
i.e as ${\cal G}\left(\frac{r}{\ell}\right)/{r^{d-2+\eta}}$. From the above results,
we find $\eta=0$, and a dynamical length $\ell$ that diverges as 
$\epsilon^{-\nu}$, with $\nu=1/4,\, 1/2$ respectively below and above the
transition. The above scaling law for $T > T_c$ is compatible with recent 
numerical simulations \cite{Glotzer,Berthier1,Berthier2}. 
It is also interesting to note that for $T=T_c$ and for small $k,\omega$ our results
imply that $\chi_{4}^{-1}$ behaves as $\Gamma k^2 + 
Z \omega^a$, which allows us to identify the dynamical exponent as $z=2/a$. 
At the critical point, length and time scales are related by an anomalous 
sub-diffusion exponent: $r \sim t^{a/2}$, with $a < 1/2$ \cite{Gotze1}.  
The above result will hold for $T > T_c$ and $\tau_f^{-1}\ll \omega
 \ll 1$, 
whereas for $\omega \tau_t \ll 1$ the denominator of the propagator will rather behave as 
$k^4 + (\tau_t \omega)^2$, corresponding to simple diffusion on 
long time scales. 
Note that the relation between $\ell$ and $\tau_t$ 
allows one to define a second 
dynamical exponent $z'=1/a+1/b=2\gamma$, echoing the presence of two diverging time scales 
for $T > T_c$. Numerically, for the Kob-Andersen LJ mixture, $z' = 4.5
\pm 0.2$ \cite{Berthier2}, a value surprisingly close to our
prediction $2 \gamma = 4.68$ \cite{Gotze1}.
However, exactly as for usual critical phenomena, one should expect 
long-wavelength fluctuations to be dominant below some upper critical dimension $d_c$ 
and changing the value of all the exponents, {\it at least sufficiently close to $T_c$}. 
The value of $d_c$ can be obtained by analyzing 
the diagrams correcting the {\sc mct} contribution to the
self-energy considered here, or using 
a Ginzburg-like criterion; both lead to $d_c=6$. The diagrammatic is in fact 
very similar to that of the $\phi^3$ theory, which is in a sense expected since the 
order parameter is here the correlation function itself, which does not have the Ising 
symmetry. More physically, one can argue that for the spatial fluctuations 
to be irrelevant, these should not blur the $\sqrt{\epsilon}$
singularity for $T< T_c$ (or the $\epsilon$ singularity
for $T > T_c$) of the non-ergodic parameter $f$ around $f_c$. 
Within a sphere of radius $\ell$, these
fluctuations are of order $\left[\ell^d \int^\ell r^{d-1} G_4(r) dr\right]^{1/2} 
\sim \ell^{(d+2)/2}$, to
be compared to the total contribution of the singularity, $\ell^d
\sqrt{\epsilon}$ for $T < T_c$ (or $\ell^d \epsilon$ for $T > T_c$).
Using $\epsilon \sim \ell^{-4}$ for $T < T_c$ ($\epsilon \sim
\ell^{-2}$ for $T > T_c$), we thus find that the fluctuations become dominant for 
large $\ell$ whenever $(d+2)/2 > d-2$, or $d < d_c=6$. A detailed calculation of the
exponents for $d < 6$ would be very interesting, to estimate how 
the mean field exponents of {\sc mct} are affected by spatial fluctuations in $d=3$. 
A naive guess, based on a percolation interpretation of the MCT
transition \cite{Biroli}, suggests $\nu \approx 0.88$ and $\eta \approx 0$. 
[The analogy with phase transitions could also shed light on the non trivial 
(fractal) structure of the mobile regions, see \cite{Weeks,Houches,Cugliandolo}.]
The knowledge of these critical exponents is crucial, since 
a quantitative fit of many experimental results has been attempted using {\sc mct} 
\cite{Gotze1}. In particular, neither the $\sqrt{\epsilon}$ singularity of 
the non-ergodic parameter for $T < T_c$ (argued to be a signature of the 
{\sc mct} singularity) nor the exponents derived above for $\chi_{4}$
are expected to remain valid for $d=3$. Nevertheless the existence of 
a diverging length scale could hint at a large (and perhaps unexpected) 
degree of universality in the dynamics of glassy systems \cite{Glotzer,Berthier2}.\\
The critical fluctuations discussed above 
should however not be confused with another type of fluctuations 
that are expected to 
destroy the {\sc mct} singularity altogether, and suppress 
the glass transition. 
These 
fluctuations are the `activated events' or `hopping processes' that 
prevent a complete
freezing of the super-cooled liquid, but 
the consistent inclusion of these in an 
extended version of {\sc mct} is still quite a challenge \cite{Gotze1,DM}. 
The way the two types of fluctuations 
interact to make any of the above finite dimensional {\sc mct} predictions 
observable is unclear to us: a second Ginzburg criterion, pertaining to 
activated fluctuations, should be devised and compared to the one above. 

The existence of a diverging 
dynamical correlation length puts strong constraints 
on any theory of glass 
forming liquids. 
We have shown that finite dimensional {\sc mct} does, at least 
qualitatively, survive the test (note that the existence of a diverging length
scale in MCT should allow for the observed decoupling between
diffusion and viscosity.) Only a more 
quantitative comparison of the predictions of {\sc mct} in three dimensions 
would allow one to 
rule it out entirely, or to confirm that it is indeed a useful picture 
(at least sufficiently far from $T_c$). Simulations of systems with
different fragilities and/or in higher dimensions 
would also be very interesting to 
gain further insight, and test different predictions or scenarii. 
For instance, the critical mobility defect scenario of \cite{Berthier2}
predicts an upper critical dimension $d_{c}=4$ with the same dynamical
exponent $z \approx 3.7$ for both strong and fragile-to-strong liquids, 
in contrast with the prediction of spatial {\sc mct}, where $d_c=6$ and 
where both $a$ and $b$ (and therefore $z$) are system-dependent.
Finally, the extension of the ideas proposed here to 
non-zero shear rates and aging situations would be extremely 
valuable (see \cite{shear,Cugliandolo}).

\acknowledgments
We thank L. Berthier, E. Bertin, S. Franz, A. Lef{\`e}vre and D. Reichmann
for useful discussions (in particular, for the last two, on the validity 
of {\sc fdt} in {\sc mct}). We are also indebted to 
S. Franz for providing us with his numerical code for
the $p=3$ spin-glass.

\end{document}